{\newcounter{numberofpages} \setcounter{numberofpages}{11} 
 
\documentclass[11pt]{article} 
\usepackage{graphicx} 
\usepackage[tbtags]{amsmath} 
\usepackage{amsthm,amsfonts,amssymb,verbatim,cite} 
\makeatletter\let\over\@@over\makeatother 
\setcounter{page}{1} 
\newcounter{finalpage} \setcounter{finalpage}{\thepage} 
\addtocounter{finalpage}{\value{numberofpages}} 
\addtocounter{finalpage}{-1} 
\typeout{****************************************************************} 
\typeout{*      Paper pages: \thepage\space --- \arabic{finalpage}      *} 
\typeout{****************************************************************} 
\thispagestyle{empty} 
\setlength{\textwidth}{5in} 
\setlength{\textheight}{8in} 
\setlength{\topmargin}{-5mm} 
\advance\baselineskip 0pt plus 1.5pt 
\advance\parskip 0pt plus 2pt 
\makeatletter\@addtoreset{equation}{section}\makeatother 
 
\newcommand{\TitAuthAffil}[3]{%
\begin{center} 
\Huge #1 
\end{center} 
\begin{center} 
{\large #2}\\ 
\vspace{0.5cm} 
{\it #3} 
\end{center} 
\vspace{1cm} 
}

\newcommand{\lb}{\left\lbrack}

\newcommand{\rb}{\right\rbrack}

\begin{document} 
 
\rightline{CLNS 98/1584} 
 
\vspace{1cm} 
 
\TitAuthAffil{On the Integrability of the Bukhvostov--Lipatov Model}%
{Marco Ameduri, Costas J.\ Efthimiou, Bogomil Gerganov}%
{Newman Laboratory of Nuclear Studies \\ 
	Cornell University \\ 
	Ithaca, NY 14853 --- USA}

\begin{abstract} 
The integrability of the Bukhvostov--Lipatov four-fermion model is  
investigated. It is shown that the classical model possesses a current of  
Lorentz spin 3, conserved both in the bulk and on the half-line for specific  
types of boundary actions. It is then established that the conservation law  
is spoiled at the quantum level --- a fact that might indicate that the  
quantum Bukhvostov--Lipatov model is not integrable, contrary to what was  
previously believed. 
\end{abstract} 
 
\section{Introduction} 
The Bukhvostov--Lipatov model (BL) is a generalization of  
the massive Thir-ring model (MTM). Correspondingly, the bosonized version  
of the  
model is a generalization of the sine-Gordon model (SG). The model was  
first introduced in a paper by Bukhvostov and Lipatov\cite{BL} in a study of  
the $O(3)$ nonlinear $\sigma$-model and has drawn recent attention in  
works by Fateev\cite{F} and Lesage et al.\ \cite{LSS}. The bosonic  
version of the model is defined by the action 
\begin{equation} 
S = \frac{1}{4\pi} \int dt dx \left[  
\frac{1}{2} {\left( \partial_\mu \phi_1 \right)}^2 +  
\frac{1}{2} {\left( \partial_\mu \phi_2 \right)}^2 + 
\lambda \cos {\beta}_1 \phi_1 \cos {\beta}_2 \phi_2 
\right] \ . 
\label{Double:cos} 
\end{equation} 
 
	It was shown in \cite{AE} that the model (\ref{Double:cos}) is not  
classically integrable, but quantum integrability has been found for  
several submanifolds in the $({\beta}_1, {\beta}_2)$-parameter  
space\cite{BL,F,LSS}.  
 
	In this work we study the fermionized model, derived from the  
Lagrangian (\ref{Double:cos}) by fermionization along the manifold proposed  
by \cite{BL}.  
We show, by explicit construction, that the fermionic BL model has a  
classically conserved current of Lorentz spin 3. The conservation of this  
current in the bulk theory is also preserved in a theory on the half-line for  
specific types of boundary actions. We then find, by using perturbed conformal 
field theory, that the classical conservation law does not survive in the  
quantum field theory, thereby suggesting that the quantum model is most  
probably not integrable, contrary to the original claim\footnote{%
	Upon completion of this work we became aware of a recent result 
	obtained by Saleur, now available in \cite{S:comm}, showing  that 
	the fermionic BL model can be made integrable using a suitable 
	regularization scheme in the Bethe Ansatz approach.%
}.

\section{Bukhvostov--Lipatov's Result} 
	In this section we summarize the main result of Bukhvostov--Lipatov's 
paper\cite{BL}. 
	Using Coleman's bosonization procedure\cite{C}, Bukhvostov and  
Lipatov have mapped the bosonic theory (\ref{Double:cos}) onto a dual  
fermionic theory. The resulting action is 
\begin{equation} 
	\begin{aligned} 
S= \frac{1}{4\pi} \int dt dx  
\hspace{-2.35pt} \left\{ \frac{{}_{}}{{}_{}} \right. 
	  & \frac{i}{2} \overline{\Psi} \gamma^\mu\partial_\mu\Psi 
          - \frac{i}{2} \partial_\mu\overline{\Psi} \gamma^\mu\Psi 
	  + \frac{i}{2} \overline{X} \gamma^\mu\partial_\mu X 
          - \frac{i}{2} \partial_\mu \overline{X} \gamma^\mu X  
\\ 
- & m \left(\overline{\Psi} \Psi + \overline{X} X\right) 
- g \left(\overline{\Psi}\gamma^\mu\Psi\right)%
    \left(\overline{X} \gamma_\mu X\right)  
\\  
- & g^\prime \left. \frac{{}_{}}{{}_{}} 
\hspace{-2mm}  
\left[{\left(\overline{\Psi}\gamma_\mu\Psi\right)}^2 +  
{\left(\overline{X} \gamma_\mu X\right)}^2 \right] \right\} \ , 
\\ 
	\end{aligned}  
\label{S:full} 
\end{equation} 
where  
\begin{equation} 
g = \pi^2\left(\frac{1}{{\beta}^2_1} - \frac{1}{{\beta}^2_2}\right)  
\ , ~~~~~~~ 
g^\prime = \frac{\pi^2}{2} \left(\frac{1}{{\beta}^2_1}+ 
\frac{1}{{\beta}^2_2}-4\right) \ . 
\label{Couplings} 
\end{equation} 
We see that the Lagrangian in (\ref{S:full}) has 2 types of four-fermion  
interactions. The term with coupling $g^\prime$ is simply the interaction of  
the MT model. The term with coupling $g$ is new and  
specific for the fermionized version of the double cosine model in  
consideration.  
 
	In their work Bukhvostov and Lipatov claimed integrability of the  
theory (\ref{S:full}) in two separate cases: 
\begin{equation} 
	\begin{align} 
1. ~~ & g = \pi^2\left(\frac{1}{{\beta}^2_1} -  
\frac{1}{{\beta}^2_2}\right) = 0 \ , 
\label{Manifold:MT} 
\\[7pt] 
2. ~~ & g^\prime = \frac{\pi^2}{2} \left(\frac{1}{{\beta}^2_1}+ 
\frac{1}{{\beta}^2_2}-4\right) = 0 \ . 
\label{Manifold:BL} 
	\end{align} 
\end{equation} 
In Case 1 (\ref{S:full}) reduces to two copies of the MT model (one for $\Psi$ 
and one for $X$), which is known to be integrable both classically\cite{MT:Cl} 
and quantum mechanically\cite{MT:Q}. In Case 2 one obtains a new  
fermionic model, to which we will refer from now on as ``the fermionic BL  
model'': 
\begin{multline} 
\hspace{-5.5pt} 
S_{BL}= \hspace{-1pt} \frac{1}{4\pi} \hspace{-3pt}\int \hspace{-4pt}dt dx  
\hspace{-2.35pt} \left\{ 
	    \frac{i}{2} \overline{\Psi} \gamma^\mu\partial_\mu\Psi 
          - \frac{i}{2} \partial_\mu\overline{\Psi} \gamma^\mu\Psi 
	 + \frac{i}{2} \overline{X} \gamma^\mu\partial_\mu X 
          - \frac{i}{2} \partial_\mu \overline{X} \gamma^\mu X  
\right. \\ 
\left. \frac{{}_{}}{{}_{}} 
\hspace{-2mm} 
- m \left(\overline{\Psi} \Psi + \overline{X} X\right) 
- g \left(\overline{\Psi}\gamma_\mu\Psi\right)%
    \left(\overline{X} \gamma^\mu X\right)  
\right\} \ .  
\label{S:BL} 
\end{multline} 
Using the Bethe Ansatz approach, Bukhvostov and Lipatov have been able to  
build the  
pseudoparticle $S$-matrix for the theory (\ref{S:BL}). They have showed  
that  
this $S$-matrix satisfies the Yang-Baxter equation for the pseudoparticle  
states. The actual physical states, however, have not been  
constructed in Bukhvostov--Lipatov's paper. To the best of our knowledge,  
computing the $S$-matrix for the physical states and, thus carrying  
out the Bethe Ansatz calculation for the model consistently to the end,  
still remains an open problem.  
 
	In the following sections we take a different point of view at the  
fermionic BL model: rather than trying to compute the $S$-matrix, we will  
try to build conserved quantities of higher tensorial rank, both in the  
classical and in the quantum version of the theory.  
 
\section{Classical Integrability} 
We will work in light-cone coordinates  
$z = \frac{1}{2} \left( t + x \right)$,  
$\bar{z} = \frac{1}{2} \left( t - x \right)$. Then we can rewrite the  
action (\ref{S:BL}) in terms of spinor components\footnote{%
	We use the following conventions: 
\begin{equation} 
   	\Psi \rightarrow \left(\begin{array}{c} \psi_{+} \\[7pt]  
\overline{\psi}_{+} \end{array}\right),  
~   	\Psi^\dagger \rightarrow \left(\psi_{-} ~ \overline{\psi}_{-} \right);  
~~	\overline{\Psi} \equiv \Psi^{\dagger} \gamma^{0}; 
~~	\gamma_0=\left[ \begin{array}{cc} 0&~ 1 \\  1&~ 0 \end{array} \right]  
\ , 
~~   	\gamma_1=\left[ \begin{array}{cc} 0&~ 1 \\ -1&~ 0 \end{array} \right] 
\nonumber 
\end{equation} 
and, similarly, for $X$.  The  
components of the metric tensor are $\lb g_{\mu\nu}\rb = \lb g^{\mu\nu} \rb =  
{\rm diag}\lb 1, -1 \rb$, so that $\gamma_{0} = \gamma^{0}$ and $\gamma_{1} =  
- \gamma^{1}$. }   
: 
\begin{equation} 
	\begin{aligned} 
	S =\ \frac{1}{4\pi} \int dt dx 
		& \left[ \frac{i}{2} \left( 
		  \psi_{+} \partial_{\bar{z}} \psi_{-} +  
		  \psi_{-} \partial_{\bar{z}} \psi_{+} + 
		  \overline{\psi}_{+} \partial_z \overline{\psi}_{-} + 
		  \overline{\psi}_{+} \partial_z \overline{\psi}_{-} 
		  \right) \right. + \\[7pt] 
	        & + \frac{i}{2} \left( 
		  \chi_{+} \partial_{\bar{z}} \chi_{-} +  
		  \chi_{-} \partial_{\bar{z}} \chi_{+} + 
		  \overline{\chi}_{+} \partial_z \overline{\chi}_{-} + 
		  \overline{\chi}_{+} \partial_z \overline{\chi}_{-} 
		  \right) + \\[7pt] 
	        & + m \left( \psi_{+} \overline{\psi}_{-} + 
		  \overline{\psi}_{+} \psi_{-} +   
			     \chi_{+} \overline{\chi}_{-} + 
		  \overline{\chi}_{+} \chi_{-} \right) - \\[7pt] 
                & - 2g \left. \left( 
		    \psi_{+}\psi_{-}\overline{\chi}_{+}\overline{\chi}_{-} 
		  + \overline{\psi}_{+}\overline{\psi}_{-}\chi_{+}\chi_{-}  
		    \right) \frac{{}_{}}{{}_{}}\hspace{-1.2mm}\right] \ . 
	\end{aligned} 
\label{L:E} 
\end{equation} 
The classical equations of motion resulting from this action are 
\begin{equation} 
	\begin{aligned} 
	\ & \pm i \partial_{\bar{z}}\psi_{\pm}(z,\bar{z})= 
		    & m \overline{\psi}_{\pm}  
		\mp & 2g \psi_{\pm}\overline{\chi}_{+}\overline{\chi}_{-}\ , \\ 
	\ & \pm i \partial_z\overline{\psi}_{\pm}(z,\bar{z})= 
		    & m \psi_{\pm}  
		\mp & 2g \overline{\psi}_{\pm}\chi_{+}\chi_{-}\ . 
	\end{aligned} 
\label{EoM} 
\end{equation} 
The corresponding set of equations for the $\chi$-fields can be obtained from  
(\ref{EoM}) with the substitution $\psi \leftrightarrow \chi$. 
 
Because of the space and time translational invariance of theory, the  
energy-momentum tensor remains conserved: 
\begin{equation} 
\partial_{\bar{z}} T_2 = \partial_z\Theta_{0} \ ,~~~~~~~~~~ 
\partial_z\overline{T}_2 = \partial_{\bar{z}} \, \overline{\Theta}_{0} \ , 
\nonumber 
\end{equation} 
where 
\begin{equation} 
	\begin{aligned} 
	T_2 &\equiv T_{zz} = i \psi_{+}\partial_z\psi_{-} +  
			     i \psi_{-}\partial_z\psi_{+} + 
			     i \chi_{+}\partial_z\chi_{-} + 
			     i \chi_{-}\partial_z\chi_{+}\ , \\ 
	\overline{T}_2 &\equiv T_{\overline{z} \, \overline{z}} = 
		i \overline{\psi}_{+}\partial_{\bar{z}}\overline{\psi}_{-} +  
		i \overline{\psi}_{-}\partial_{\bar{z}}\overline{\psi}_{+} +  
		i \overline{\chi}_{+}\partial_{\bar{z}}\overline{\chi}_{-} +  
		i \overline{\chi}_{-}\partial_{\bar{z}}\overline{\chi}_{+}\ ,  
\\ 
	-\Theta_0 = -\overline{\Theta}_0 &\equiv T_{z \, \overline{z}} =  
	T_{\overline{z} \, z} = 
		- m \left(  
		    \psi_{+}\overline{\psi}_{-} + \overline{\psi}_{+}\psi_{-} + 
		    \chi_{+}\overline{\chi}_{-} + \overline{\chi}_{+}\chi_{-} 
		    \right). 
	\end{aligned} 
\label{En:Mom} 
\end{equation} 
 
	The existence of integrals of motion of higher Lorentz spin is  
considered to be a strong indication  
for the classical integrability of the theory. We have been able to show that  
the fermionic BL model (\ref{L:E}) has a classically conserved charge of spin  
3 in the bulk: 
\begin{equation} 
Q_3 = \int_{-\infty}^{+\infty} dx \left( T_4 - \Theta_2 \right)  
~~~ {\rm and} ~~~  
\overline{Q}_3 = \int_{-\infty}^{+\infty} dx \left( \overline{T}_4 - 
\overline{\Theta}_2 \right) \ ,  
\label{Q3} 
\end{equation} 
where the densities $T_4$ and $\Theta_2$ are given by: 
\begin{equation} 
	\begin{aligned} 
	T_{4} = \ & -i \psi_{+}\partial_z^3\psi_{-} +  
		  6g \partial_z\psi_{+}\partial_z\psi_{-}\chi_{+}\chi_{-} +  
		  \ ({\scriptstyle +} \leftrightarrow {\scriptstyle -}) +  
		  \ (\psi \leftrightarrow \chi) 
	\end{aligned} 
\label{T4} 
\end{equation} 
and 
\begin{equation} 
	\begin{aligned} 
	-m^{-1} \Theta_{2} = \ 
	& \overline{\psi}_{+}\partial_z^2\psi_{-} + 
          \partial_z^2\psi_{+}\overline{\psi}_{-} + 
	  2ig \overline{\psi}_{+}\partial_z\psi_{-}\chi_{+}\chi_{-} -  
	  2ig \partial_z\psi_{+}\overline{\psi}_{-}\chi_{+}\chi_{-} +	 
			\\ 
	& + 4ig \overline{\psi}_{+}\psi_{-}\chi_{+}\partial_z\chi_{-} - 
	  4ig \psi_{+}\overline{\psi}_{-}\partial_z\chi_{+}\chi_{-} + 
	  4ig \overline{\psi}_{+}\psi_{-}\partial_z\chi_{+}\chi_{-} - 
			\\ 
	& - 4ig \psi_{+}\overline{\psi}_{-}\chi_{+}\partial_z\chi_{-} +  
	    8mg \psi_{+}\psi_{-}\chi_{+}\chi_{-} +		  
	    \ (\psi \leftrightarrow \chi) \ , 
 	\end{aligned} 
\label{Theta2} 
\end{equation} 
along with analogous expressions for $\overline{T}_4$ and 
$\overline{\Theta}_2$. These  
quantities satisfy the conservation equations 
\begin{equation} 
	\partial_{\bar{z}} T_4 = \partial_z\Theta_2,~~~~~~~~~~ 
	\partial_z\overline{T}_4 = \partial_{\bar{z}}\overline{\Theta}_2  
\ . 
\label{ClassCL} 
\end{equation} 
The conserved current above is peculiar to the BL model. It naturally reduces  
to the classically conserved spin 3 current of the MT model in the limit  
$X=\Psi$ (cf. \cite{IKZ}). We intend to generalize our result to conserved  
quantities of arbitrary spin by using the methods, developed in Refs.  
\cite{MT:Cl,GGKM:KdV,Lax,TF,GSc,FGSc} for the classical MT, 
Korteweg-de Vries, and SG models. 
 
We have also found that the more general 2-fermion model (\ref{S:full})  
does not possess a classically conserved spin 3 current for arbitrary values  
of the couplings $g$ and $g^\prime$. A conservation law of spin 3  
holds only in the special cases $g=0$ (2$\times$MT model) and $g^\prime=0$ 
(BL model). Therefore, at the classical level, the model (\ref{S:full})  
is integrable precisely in the cases (\ref{Manifold:MT}) and  
(\ref{Manifold:BL}), suggested by Bukhvostov and Lipatov. 
 
\subsection{Conserved Quantities of Higher Spin in the Presence of a Boundary} 
 
	It is interesting to consider the BL theory on the half-line. The  
action is modified as follows: 
\begin{equation} 
S =  
\int_{-\infty}^{+\infty} dt \int_{-\infty}^0 dx \mathcal{L} + 
\int_{-\infty}^{+\infty} dt \mathcal{B} \ , 
\label{S:bound} 
\end{equation} 
where $\mathcal{B}$ is a boundary potential. 
It can contain operators built out of bulk fields, evaluated at the  
boundary, $x=0$, as well as new boundary degrees of freedom.  
 
	A conserved quantity on $\mathbb{R}_x$ is not necessarily conserved  
on the half-line. Indeed, if there exists a local conservation law of spin $s$, 
$\partial_{\bar{z}} T_{s+1} = \partial_z\Theta_{s-1}$, 
$\partial_{\bar{z}} \overline{T}_{s+1} = \partial_z\overline{\Theta}_{s-1}$,  
in the theory on the half-line we have 
\begin{equation} 
	\begin{aligned} 
\frac{d}{d t} 
\left( 
Q_s + \overline{Q}_s  
\right)  
= \ & \frac{d}{d t}\int_{-\infty}^0 dx  
\left(  
T_{s+1} + \Theta_{s-1} + \overline{T}_{s+1} + \overline{\Theta}_{s-1}  
\right) = \\ 
= \ &  
\left(  
\overline{T}_{s+1} + \overline{\Theta}_{s-1} - T_{s+1} - \Theta_{s-1}  
\right)_{x=0} 
	\end{aligned} 
\label{Cons:bound} 
\end{equation} 
which, in general, is different from 0. If, however, the RHS of  
(\ref{Cons:bound})  
can be written as a total time-derivative of some function $F_s(t)$ then the  
charge $P_s \equiv \left( Q_s + \overline{Q}_s \right) - F_s$ will be  
conserved on the half-line. Whether a conservation law of higher spin survives 
in the boundary theory is entirely dependent on the form of $\mathcal{B}$. 
Therefore, finding boundary potentials for which such function $F_s$ exists
provides a method for studying the classical integrability of boundary field 
theories \cite{IKZ}.
 
	Using the above technique, we have been able to show that the 
conservation of the BL spin 3 current (\ref{Q3}) is preserved on the half-line 
for several types of boundary potentials. A list of such boundary potentials 
is provided in the Appendix. Similar methods have been applied to study the 
integrability on the half-line of the super--Liouville theory \cite{Pr} and, 
very recently, of the $O(N)$ nonlinear $\sigma$-model and the $O(N)$ 
Gross-Neveu model \cite{MdM}.

	An extensive discussion of quantum integrability in the presence of a 
boundary and methods for computing the boundary $S$-matrix can be found 
in \cite{GZ}.

\section{Quantum Integrability} 
 
	In this section we will study the modifications to the classical  
conservation law (\ref{ClassCL}) due to quantum corrections. A powerful tool  
for building conserved quantities for 2D quantum models is the technique of  
perturbed conformal field theory\cite{CPT}. By treating a 2D QFT as a  
perturbed CFT, 
it is possible to study which, if any, of the infinite conservation laws 
present in any CFT  
survive the perturbation. Zamolodchikov's paper\cite{CPT} also provides us  
with an easy way for computing the conserved current densities explicitly. 
 
	There are several difficulties in applying the formalism of perturbed  
CFT directly to the fermionic model (\ref{S:BL}). In principle, one could  
regard the model as perturbed free massless fermion theory, treating both the  
$m$- and the $g$-terms as perturbations. In order to  
discover non-trivial corrections to $T_4$, however, one needs to go to at  
least second order in PT where the simple Zamolodchikov's formula  
for computing $\partial_{\bar{z}} T_4$ is no longer valid. Another problem of  
this approach would be the fact that the $g$-term is a marginal  
operator and Zamolodchikov's counting argument\cite[p.650]{CPT} does not  
apply --- the perturbation series in $g$ is, in general, infinite. 
 
	We will, therefore, rebosonize the fermionic current (\ref{T4}) to 
study its quantum conservation in the double cosine model (\ref{Double:cos}), 
treating the term $\lambda \cos{\beta}_1\phi_1\cos{\beta}_2\phi_2$  
as a single relevant perturbation to the conformal theory of massless free  
bosons.  
 
	We find that, up to some coefficients to be determined later, 
\begin{equation} 
	\begin{aligned} 
\psi_{+}\partial^3_z\psi_{-} + \psi_{-}\partial^3_z\psi_{+} ~~  
& \propto ~~{\left( \partial^2_z \vartheta_1 \right)}^2 +  
		    {\left( \partial_z \vartheta_1 \right)}^4 \\[7pt] 
\partial_z\psi_{+}\partial_z\psi_{-}\chi_{+}\chi_{-} ~~ 
& \propto ~~\partial^3_z\vartheta_1\partial_z\vartheta_2 + 
	{\left( \partial_z \vartheta_1 \right)}^3\partial_z\vartheta_2 
	\end{aligned} 
\label{Boson:Ops} 
\end{equation} 
and, similarly, for the $\left( \psi \leftrightarrow \chi \right)$-terms in 
$T_4$. 
 
	Finally, let's note that quantum corrections will, in principle,  
modify the coefficients of the classical $T_4$. We therefore prefer to leave  
them as arbitrary functions of $g$ and then fix them while computing the  
conserved current via perturbed CFT. 
 
	Using the bosonization operator identities (\ref{Boson:Ops}), we can  
write $T_4$ in terms of the boson fields  
$\phi_1$ and $\phi_2$%
\footnote{The fields $\phi_1$ and $\phi_2$ are linear combinations of  
$\vartheta_1$ and $\vartheta_2$}%
: 
\begin{equation} 
	\begin{aligned} 
T_4 =\ &a_1 {\left( \partial^2_z \phi_1 \right)}^2 + \ 
      	a_2 {\left( \partial^2_z \phi_2 \right)}^2 + \ 
 	b_1 {\left( \partial_z \phi_1 \right)}^4 + \ 
	b_2 {\left( \partial_z \phi_2 \right)}^4 + \\[7pt] 
  + \ &	c    {\left( \partial_z \phi_1 \right)}^2 
	     {\left( \partial_z \phi_2 \right)}^2 + \ 
      	d \  \partial^2_z\phi_1\partial^2_z\phi_2 + \\[7pt] 
  + \ &	f_1 {\left( \partial_z \phi_1 \right)}^3\partial_z\phi_2 + \ 
      	f_2 \partial_z\phi_1{\left( \partial_z \phi_2 \right)}^3 + \\[7pt] 
  + \ &	h_1 {\left( \partial_z \phi_1 \right)}^2\partial^2_z\phi_2 + \ 
      	h_2 \partial^2_z\phi_1{\left( \partial_z \phi_2 \right)}^2 \ . 
	\end{aligned} 
\label{T4:bosonic} 
\end{equation} 
The expression above is, in fact, the most general Ansatz for $T_4$ for 
the double cosine model. It includes  
all\footnote{There are some other operators of mass dimension 4 that could be 
	     included but all of them are identical to the operators in  
	     (\ref{T4:bosonic}) up to total $\partial_z$-derivatives.}  
operators of mass dimension 4 
with arbitrary coefficients. These coefficients are functions of  
$({\beta}_1, {\beta}_2)$ or, via (\ref{Couplings}), of $g$. As it  
turns out, the requirement that $T_4$ be conserved in the perturbed CFT is  
very restrictive and gives enough information to compute the exact form of  
these functions. 
 
	In CFT $T_4$ is a holomorphic function and $\partial_{\bar{z}}T_4 = 0$. 
In the perturbed QFT that is no longer true and we can compute  
$\partial_{\bar{z}}T_4$, using Zamolodchikov's formula\cite[{eq.(3.14)}]{CPT}: 
\begin{equation} 
\partial_{\bar{z}}T_4 = \lambda \oint_z \frac{d\zeta}{2\pi i}  
\cos{\beta}_1\phi_1(\zeta,\bar{z}) \cos{\beta}_2\phi_2(\zeta,\bar{z}) 
T_4(z) 
\label{Zamo} 
\end{equation} 
If the RHS of (\ref{Zamo}) can be expressed as a total $\partial_z$-derivative 
of some local operator, $\partial_z \Theta_2$, the conservation law of spin 3  
survives in the perturbed QFT and has the form (\ref{ClassCL}), $T_4$ and 
$\Theta_2$ being now the quantum conserved densities.  
 
	In starting this calculation, our goal was to find all the  
conditions on the couplings ${\beta}_1$ and ${\beta}_2$ for which the  
spin 3 charge is conserved. We expected to find the `BL manifold'  
(\ref{Manifold:BL}) as one of the integrable cases and then, by fermionizing  
back, to obtain an exact quantum expression for the spin 3 conserved current 
of the fermionic BL model. 
 
	As a result of the calculation one finds that the spin 3 current is 
conserved only in 3 cases: 
\begin{equation} 
	\begin{align} 
{\beta}^2_1 - {\beta}^2_2 = \ & 0 \ , 
\label{2sG} \\ 
{\beta}^2_1 + {\beta}^2_2 = \ & 1 \ , 
\label{Saleur1} \\ 
{\beta}^2_1 + {\beta}^2_2 = \ & 2 \ . 
\label{Saleur2} 
	\end{align} 
\end{equation} 
The first manifold is trivial: when ${\beta}^2_1 = {\beta}^2_2$ the  
double cosine model decouples into 2 sine-Gordon models and, of course, is  
integrable both classically and quantum mechanically. These manifolds  
have been previously identified by Fateev\cite{F} and by Lesage et al. 
\cite{LSS}. On the BL manifold (\ref{Manifold:BL}) the charge  
$Q_3 = \int dx \left( T_4 - \Theta_2 \right)$ is not conserved, except  
in the trivial case when the manifolds (\ref{2sG}), (\ref{Manifold:BL}), and  
(\ref{Saleur1}) intersect each other (free fermion point). Therefore, we  
conclude that the spin 3 conservation law of the fermionic BL theory is  
spoiled by the quantum corrections. 
 
	The above result becomes even more clear in fermionic language. Let's  
look at the $(g, g^\prime)$-parameter space, where $g$ and $g^\prime$ are the  
couplings of the general 2-fermion action (\ref{S:full}). As we showed at the  
end of Section 3.1, the general model has a classically conserved charge  
of spin 3 only if either $g=0$ or $g^\prime=0$. Therefore, the classical  
integrable manifolds are simply the axes of the  $(g, g^\prime)$-plane, $g=0$ 
and $g^\prime=0$. 
 
	To find the quantum integrable manifolds, we simply need to  
rewrite equations (\ref{2sG})--(\ref{Saleur2}) in terms of $g$ and $g^\prime$  
via relations (\ref{Couplings}). 
\begin{equation} 
	\begin{align} 
{\beta}^2_1 - {\beta}^2_2 = \ 0 \ & \longrightarrow ~~ \ g = 0 \ , 
\label{QMT} \\ 
{\beta}^2_1 + {\beta}^2_2 = \ 1 \ & \longrightarrow ~~  
	{\left(\frac{2g^\prime}{\pi^2}+2\right)}^2 -  
	{\left(\frac{g}{\pi^2}\right)}^2 = 4 \ , 
\label{QFM} \\ 
{\beta}^2_1 + {\beta}^2_2 = \ 2 \ & \longrightarrow ~~ 
	{\left(\frac{2g^\prime}{\pi^2}+3\right)}^2 -  
	{\left(\frac{g}{\pi^2}\right)}^2 = 1 \ . 
\nonumber 
	\end{align} 
\end{equation} 
\vspace{-1cm}
\begin{figure}[htb] 
\hspace{11mm} \includegraphics[width=8cm]{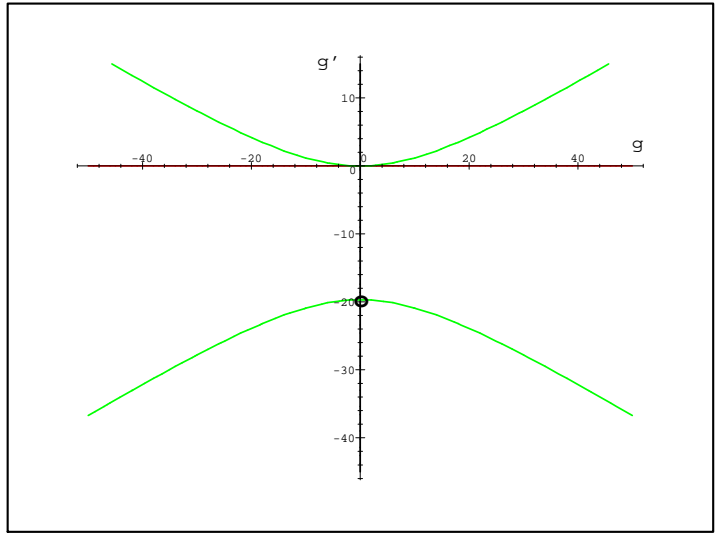} 
\vspace{-2cm} 
\caption{Fermionic parameter space.} 
\label{Par-ferm} 
\end{figure} 

 	We see that the manifold $g=0$ is present both in the classical and 
in the quantum case. This merely reflects the fact that massive Thirring
model is integrable both classically and quantum mechanically. In contrast, 
the fermionic BL model (\ref{S:BL}), obtained by setting $g^\prime=0$ in the 
general action (\ref{S:full}), has a charge of Lorentz spin 3 which is 
conserved classically but not quantum mechanically.
 
\section{Conclusion} 
 
We have shown that the fermionic Bukhvostov--Lipatov model, given by the  
action  (\ref{S:BL}), admits a nontrivial classical integral of motion of  
spin 3, both in the bulk and for specific types of boundary actions in the  
theory on the half line. This conservation law holds quantum mechanically  
only at the free  
fermion point $g=0$ and is spoiled by quantum corrections for generic values  
of the coupling $g$. The more general fermionic model (\ref{S:full}) admits a  
quantum conservation law of spin 3 for the specific relation (\ref{QFM})  
between the couplings $g$ and $g^\prime$. The study of the spin 3 conservation 
laws, therefore, suggests that the integrable manifold $(g$ free,  
$g^\prime=0)$ proposed by Bukhvostov and Lipatov does not survive 
in the quantum field theory.  

\section*{Acknowledgements} 
 
	We would like to thank Andr\'{e} LeClair for support and for helping 
us with many useful discussions and insights during the completion of this 
work, Fr\'ed\'eric Lesage for a discussion on the double-cosine model, and
Hubert Saleur for sharing with us his results. 
 
\section*{Appendix} 
 
	The following is a list of boundary potentials for which the  
conservation of the spin 3 charge of the classical fermionic BL  
model is preserved in the theory on the half-line. This list does not  
claim to be exhaustive. 
 
	If one leaves out all additional boundary degrees of freedom, and  
considers the boundary actions which are functionals of the bulk fields only, 
the most general Ansatz for a boundary potential is: 
\begin{equation} 
	\begin{aligned} 
	{\cal B}(\psi_{+}, \ & \psi_{-}, \overline{\psi}_{+},  
\overline{\psi}_{-}; \chi_{+}, \chi_{-}, \overline{\chi}_{+},   
\overline{\chi}_{-}) =  
\\ 
	\ = \ & 
	a_1 \psi_{-} \psi_{+} + a_2 \overline{\psi}_{-} \overline{\psi}_{+} + 
  	b_1 \chi_{-} \chi_{+} + b_2 \overline{\chi}_{-} \overline{\chi}_{+} + 
\\ 
	\ &+ 
i a_3 \lb \psi_{-} \overline{\psi}_{-} - \overline{\psi}_{+} \psi_{+} \rb + 
i a_4 \lb e^{i\alpha} \psi_{-} \overline{\psi}_{+} - e^{-i\alpha} 
\overline{\psi}_{-} \psi_{+} \rb + 
\\ 
	\ &+  
i b_3 \lb \chi_{-} \overline{\chi}_{-} - \overline{\chi}_{+} \chi_{+} \rb + 
i b_4 \lb e^{i\beta} \chi_{-} \overline{\chi}_{+} - e^{-i\beta} 
\overline{\chi}_{-} \chi_{+} \rb + 
\\ 
	\ &+  
   i c_1 \lb e^{i\gamma_1} \psi_{-} \chi_{-} -  
e^{-i\gamma_1} \chi_{+} \psi_{+} \rb + 
   i c_2 \lb e^{i\gamma_2} \psi_{-} \chi_{+} -  
e^{-i\gamma_2}\chi_{-} \psi_{+} \rb + 
\\ 
	\ &+  
   i d_1 \lb e^{i\delta_1} \overline{\psi}_{-} \overline{\chi}_{-} -  
e^{-i\delta_1} \overline{\chi}_{+} \overline{\psi}_{+} \rb + 
   i d_2 \lb e^{i\delta_2} \overline{\psi}_{-} \overline{\chi}_{+} -  
e^{-i\delta_2} \overline{\chi}_{-} \overline{\psi}_{+} \rb + 
\\ 
	\ &+  
   i f_1 \lb e^{i\varphi_1} \psi_{-} \overline{\chi}_{-} -  
e^{-i\varphi_1} \overline{\chi}_{+} \psi_{+} \rb + 
   i f_2 \lb e^{i\varphi_2} \psi_{-} \overline{\chi}_{+} -  
e^{-i\varphi_2} \overline{\chi}_{-} \psi_{+} \rb + 
\\ 
	\ &+ 
   i f_3 \lb e^{i\varphi_3} \overline{\psi}_{-} \chi_{-} -  
e^{-i\varphi_3} \chi_{+} \overline{\psi}_{+} \rb + 
   i f_4 \lb e^{i\varphi_4} \overline{\psi}_{-} \chi_{+} -  
e^{-i\varphi_4} \chi_{-} \overline{\psi}_{+} \rb \ . 
	\end{aligned} 
\nonumber 
\end{equation} 
This Ansatz gives 8 linear equations of motion for the bulk fields at the  
boundary, depending on 26 real parameters, 16 amplitudes and 10 phases.  
In order to have non-trivial solution to this linear system we require that  
the $8 \times 8$ matrix of coefficients be of rank 6 or smaller. Because  
of the size of the matrix it is difficult to study the problem in all its  
generality, but we list here some interesting particular cases: 
	 
	1. No terms mixing the $\Psi$- and the $X$-fields appear in the  
boundary action, i.e. the coefficients $c_1$, $c_2$, $d_1$, $d_2$, $f_1$,  
$f_2$, $f_3$, and $f_4$ are equal to $0$. In this case the integrable boundary  
actions are linear combinations of the integral boundary actions for the MT  
model \cite{IKZ}: 
 
\begin{center} 
\begin{tabular}{|c|c|c|c|c|c|c|c|} 
\hline 
$a_1$ & $a_2$ & $a_3$ & $a_4$ & $b_1$ & $b_2$ & $b_3$ & $b_4$ \\  
\hline  
\hline 
\ & \ & \ & \ & \ & \ & \ & \ \\[-7pt] 
    $\frac{\tan\kappa_0}{2}$ & $-\frac{\tan\kappa_0}{2}$ & 
    $\frac{1}{2\cos\kappa_0}$ & $0$ & 
    $\frac{\tan\mu_0}{2}$ & $-\frac{\tan\mu_0}{2}$ & 
    $\frac{1}{2\cos\mu_0}$ & $0$ \\[6pt] 
    $\frac{\tan\kappa_0}{2}$ & $-\frac{\tan\kappa_0}{2}$ & 
    $\frac{1}{2\cos\kappa_0}$ & $0$ &  
    $\frac{\tan\nu_0}{2}$ & $\frac{\tan\nu_0}{2}$ & 
    $0$ & $\frac{1}{2\cos\nu_0}$ \\[6pt] 
    $\frac{\tan\lambda_0}{2}$ & $\frac{\tan\lambda_0}{2}$ & 
    $0$ & $\frac{1}{2\cos\lambda_0}$ & 
    $\frac{\tan\mu_0}{2}$ & $-\frac{\tan\mu_0}{2}$ & 
    $\frac{1}{2\cos\mu_0}$ & $0$ \\[6pt] 
    $\frac{\tan\lambda_0}{2}$ & $\frac{\tan\lambda_0}{2}$ & 
    $0$ & $\frac{1}{2\cos\lambda_0}$ & 
    $\frac{\tan\nu_0}{2}$ & $\frac{\tan\nu_0}{2}$ & 
    $0$ & $\frac{1}{2\cos\nu_0}$ \\[6pt] 
\hline 
\end{tabular}  
\end{center} 
where $\kappa_0$, $\lambda_0$, $\mu_0$, and $\nu_0$ are free real parameters 
$\neq \frac{k\pi}{2}$.  
 
	2. Only terms mixing the $\Psi$- and the $X$-fields are present. 
$a_1$, ..., $a_4$ and $b_1$, ..., $b_4$ are equal to $0$. If, in addition, we  
consider the even simpler sub-case when only terms mixing $\psi_{\pm}$ with  
$\overline{\chi}_{\pm}$ and $\overline{\psi}_{\pm}$ with $\chi_{\pm}$ are  
present, i.e. also $c_1$, $c_2$, $d_1$, $d_2$ vanish, the integrable boundary  
actions are given by:  
  
\begin{center} 
\begin{tabular}{|c|c|c|c|} 
\hline 
$f_1$ & $f_2$ & $f_3$ & $f_4$ \\  
\hline  
\hline 
\ & \ & \ & \ \\[-7pt] 
    $\frac{1}{2} \pm p_0$ & $p_0$ & $\frac{1}{2} \pm q_0$ & $q_0$ \\[6pt] 
    $\pm \frac{1}{2}$ & $0$ & $r_0$ & $s_0$ \\[6pt] 
    $0$ & $\pm \frac{1}{2}$ & $r_0$ & $s_0$ \\[6pt] 
    $u_0$ & $v_0$ & $\pm \frac{1}{2}$ & $0$ \\[6pt] 
    $u_0$ & $v_0$ & $0$ & $\pm \frac{1}{2}$ \\[6pt] 
\hline 
\end{tabular}  
\end{center} 
where $p_0$, $q_0$, $r_0$, $s_0$, $u_0$, and $v_0$ are free real parameters.  
The integrable boundary actions of this type are specific for the boundary BL  
model.


\end{document}